\documentclass[twocolumn,amsmath,superscriptaddress,amssymb,amsfonts,aps,prb,floatfix,10pt]{revtex4-1}
\usepackage{graphicx, color,rotating} 
\usepackage{dcolumn} 
\usepackage{bm} 
\usepackage{epsfig}
\usepackage{soul}
\usepackage{subfigure}
\usepackage{etoolbox}

\usepackage{graphicx}
\usepackage[usenames]{xcolor}
\usepackage{printlen}
\definecolor{KITgreen}  {RGB}{  0,150,130} 
\definecolor{KITgreen70}{RGB}{ 76,181,167} 
\definecolor{KITgreen50}{RGB}{127,202,192} 
\definecolor{KITgreen30}{RGB}{178,223,217} 
\definecolor{KITgreen15}{RGB}{217,239,236} 

  \definecolor{KITblue}  {cmyk}{ 88,55,0,0} 
  \definecolor{KITblue70}{cmyk}{ 56,33,0,0} 
  \definecolor{KITblue50}{cmyk}{ 40,25,0,0} 
  \definecolor{KITblue30}{cmyk}{ 24,15,0,0} 
  \definecolor{KITblue15}{cmyk}{ 12,7.5,0,0} 

  \definecolor{KITblack}  {RGB}{  0,  0,  0} 
  \definecolor{KITblack70}{RGB}{ 77, 77, 77} 
  \definecolor{KITblack50}{RGB}{128,128,128} 
  \definecolor{KITblack30}{RGB}{179,179,179} 
  \definecolor{KITblack15}{RGB}{217,217,217} 

  \definecolor{KITpalegreen}{RGB}{130,190,60}

  \definecolor{KITyellow}{RGB}{250,230,20}

  \definecolor{KITorange}{RGB}{220,160,30}

  \definecolor{KITbrown}{RGB}{160,130,50}

  \definecolor{KITred}{RGB}{160,30,40}

  \definecolor{KITlila}{RGB}{160,0,120}

  \definecolor{KITcyanblue}{RGB}{80,170,230}

\usepackage{colortbl}

\usepackage[plainpages=false, pdfpagelabels, bookmarksopen, pdfborder={0 0 0},colorlinks=true,
urlcolor=black,citecolor=KITblue,linkcolor=KITred]{hyperref}

\usepackage{tikz}
\usetikzlibrary{arrows,decorations.pathmorphing,decorations.markings,backgrounds,positioning,fit,trees}
\tikzset{->-/.style={decoration={
  markings,
  mark=at position #1 with {\arrow{>}}},postaction={decorate}}}
\tikzset{
   vertex/.style={circle, inner sep=0pt, minimum size=5pt,fill=black,label=#1},
vertex/.default=\text{},
   crossing/.style={circle, inner sep=0, minimum size=0,label=#1},
crossing/.default=\text{},
   named/.style={draw,circle, inner sep=0pt, minimum size=12pt},
   namedE/.style={draw,circle, inner sep=0pt, minimum size=7pt},
   bath/.style={draw,thick,->-=.5}, bath/.default={},
   bathr/.style={draw,thick,dashed,->-=.5},
   bathNA/.style={draw,thick}, bath/.default={},
   bathrNA/.style={draw,thick,dashed},
   time/.style={draw,dashed,thin},
   timeAxes/.style={draw,thick,->-=1}, timeAxes/.default={},
   system/.style={draw,thick}, system/.default={},
   syslabel/.style={midway,auto,green!40!black}}
\tikzset{every picture/.append style={node distance=1.5*1cm and 1.5*2cm,bend
angle=60}}
\DeclareMathOperator{\Tr}{Tr}

\newcommand{\bra}[1]{\langle #1|}
\newcommand{\ket}[1]{|#1\rangle}

\newcommand{\pd}{{\phantom\dagger}}
\renewcommand{\Im}{\operatorname{Im}}
\renewcommand{\Re}{\operatorname{Re}}

\begin{document}
\title{Analyzing the spectral density of a perturbed analog  quantum simulator using Keldysh formalism}

\author{Sebastian Zanker}
\affiliation{Institut f\"ur Theoretische Festk\"orperphysik, Karlsruhe Institute of Technology, D-76128 Karlsruhe,
Germany}

\author{Iris Schwenk}
\affiliation{Institut f\"ur Theoretische Festk\"orperphysik, Karlsruhe Institute of Technology, D-76128 Karlsruhe,
Germany}

\author{Jan-Michael Reiner}
\affiliation{Institut f\"ur Theoretische Festk\"orperphysik, Karlsruhe Institute of Technology, D-76128 Karlsruhe,
Germany}

\author{Juha Lepp\"akangas}
\affiliation{Institut f\"ur Theoretische Festk\"orperphysik, Karlsruhe Institute of Technology, D-76128 Karlsruhe,
Germany}

\author{Michael Marthaler}
\affiliation{Institut f\"ur Theoretische Festk\"orperphysik, Karlsruhe Institute of Technology, D-76131 Karlsruhe,
Germany}


\date{\today}

\begin{abstract}
Simulation of interacting electron systems is one of the great challenges of modern quantum chemistry and solid state physics.
Controllable quantum systems offer the opportunity to create  artificial structures
which mimic the system of interest.
An interesting quantity to extract from these quantum simulations is the spectral function.
We map a noisy quantum simulator onto a fermionic system and investigate the influence of decoherence on the simulation of the spectral density using a diagrammatic approach on Keldysh contour.
We show that features stronger than the single-qubit decoherence rate can be resolved, while weaker features wash out.
For small systems, we compare our Keldysh approach to master-equation calculations. 
\end{abstract}

\maketitle
\section{Introduction}
Using quantum mechanics
we can describe many interesting systems in a wide range of fields, covering biology, chemistry, and (solid state) physics.
The interacting many-body Schr\"odinger equation provides a microscopic description of atoms, molecules, and solids.
However, due to the exponential growth of the Hilbert space with the system size, the exact solution of the full interacting Schr\"odinger equation becomes numerically demanding even for systems of moderate size.\cite{TRO15}
This problem is inherent to the simulation of quantum systems on conventional computers, limiting their application to approximations of the full Schr\"odinger equation.
For example, density functional theory (DFT)\cite{KOH64} has been used with great success to calculate electronic properties of solids and molecules on classical computers for the past decades. 
However, DFT, as other approximation schemes, is limited to weakly interacting systems and many interesting systems such as high-T$_\text{C}$ superconductivity\cite{BED86} or transition metals,\cite{JIA12,CRA09} are beyond the scope of DFT. 
The idea to use well-controllable quantum systems to overcome these difficulties\cite{Feynman1982} promises an exponential speedup compared to classical numerics.\cite{Loyd1996,Zalka1998}
Quantum simulation\cite{Johnson2014,FRA09,ZOL12} on different physical systems, such as trapped ions,\cite{BLA12} ultra cold gases,\cite{Jaksch2005,BLO12} or superconducting circuits,\cite{Shnirman2001,DS13,Houck2012,BLK+15,HWS14} have been proposed to solve problems in different fields, such as quantum chemistry \cite{WBA11,KWP+11} and strongly-correlated electrons.\cite{BLK+15,WHW+15,Troyer16}
The idea behind quantum simulation is to map the Hamiltonian onto a controllable quantum system -- the quantum simulator -- that mimics the simulated system.

One fundamental roadblock on the way to feasible quantum simulation is the  limited coherence time of qubits.\cite{MSS05,DS13}
Universal digital quantum simulation by discretization of the time evolution relies on the application of gates.\cite{BLK+15,HWS14,Lanyon2011,BAR16}
While quantum error correction is possible,\cite{MAR12,KNI97} the number of coherently applied gates needed to calculate molecules of relevant size exceeds the capability of quantum hardware available in the near future.\cite{TRO14}
Analog quantum simulation is a promising alternative for first applications for quantum simulation.\cite{MAN02,Sarovar2017}
Many proposals for analog quantum simulation of spin systems,\cite{Porras2004,Monroe2010,Greif2013,Schaetz2008,Greiner2011}, Hubbard\cite{JAN,Russell2015} and Holstein\cite{Tian2013} models, exciton transport,\cite{Aspuru2004} and many other models exist or have been realized on different plattforms already.\cite{Wenz2013,Bloch2012,ZOH13}

At present it remains unclear how to analyze the effect of decoherence in large systems of coupled qubits forming an analog quantum simulator.\cite{Hauke2012, Iris16}
In this article, we demonstrate the application of established many-body physics methods on systems of many qubits
to analyze the coherence behavior of a quantum simulator consisting of many fault-prone qubits.
An important quantity of interest we would like to extract from a quantum simulator is the spectral function,
which can also be formulated in terms of the Green's function $G(\omega)$.
We expect intuitively that decoherence rates of single qubits determine the spectral resolution of a quantum simulator.
Using Wick's theorem and many-body perturbation theory, we formulate  a connection between the ideal Green's function $G_0$ and the simulated, decoherence afflicted perturbed Green's function $G$:
\begin{equation}\label{eq:Central_Result_in_Intro}
 G(\omega)=(G_0^{-1}(\omega)-\Sigma(\omega))^{-1} \, .
\end{equation}
The self-energy $\Sigma$ describes the influence of the environment on the quantum simulator and is defined as the sum of all irreducible diagrams.
Using this connection we investigate how single-qubit decoherence affects the spectral function of several multi-qubit systems,
in and out of equilibrium.

We study systems of qubits which can be mapped to a fermionic system via a Jordan-Wigner transformation.\cite{JW28}
In Sec.~\ref{sec:GeneralDiscussion}, we discuss both directions of this mapping.
We first introduce the Jordan-Wigner transformation to map a fermionic system on a quantum simulator. 
After this we show how to map the coupling of qubits to the environment onto fermions using the same transformation. 
With this transformation we can map the noisy quantum simulator to a system of fermions coupled to a bath, similar to electron-phonon coupling.
This system can be tackled with standard many-body perturbation theory, introduced in Sec.~\ref{sec:KeldyshContour}.
There, we use the transformation to fermionic systems to obtain a diagrammatic expansion in the coupling to the environment on Keldysh contour.
In particular, we derive Eq.~\eqref{eq:Central_Result_in_Intro}. 
In Sec.~\ref{sec:SteadyState}, we apply this method to a simple chain of qubits with dephasing due to a bosonic bath and decay due to two-level systems.
We compare our results for the steady-state properties with results obtained by master-equation methods.
We show that we can  qualitatively understand the decoherence of large systems with our method. 
In Sec.~\ref{sec:time_dependence}, we show that our method can also be used to model transient evolution of a chain of qubits
after an initialization into a non-thermal state. Additionally, we compare our results with corresponding master-equation simulations.
Conclusions are given in Sec.~\ref{sec:Conclusions}.

\section{General Discussion}\label{sec:GeneralDiscussion}
In this section, we first describe the mapping of a fermionic system to a quantum simulator consisting of spin-$\frac12$ qubits. 
Afterwards, we use this method to  map the Hamiltonian of a noisy quantum simulator back to a fermionic system
which we need to be able to apply many-body perturbation theory.

\subsection{Fermionic system mapped to qubits}
In second quantization, the Hamiltonian $H_f$ of an electronic system depends on the creation (annihilation) operator 
$c_i^\dagger$ ($c_i$) of an electron in state $i$, where $i$ contains information about site, orbital, and spin. 
To simulate such a Hamiltonian on an analog quantum simulator consisting of coupled qubits, we have to map the fermionic operators to a spin-$\frac12$ system that describes the physical qubits. 
This physical difference reflects into different commutation relations for spin-$\frac12$ Pauli operators $\hat\sigma_{\pm}$ and fermionic operators.\cite{WBA11} While the latter always anti-commute
\begin{equation}
 \{c_i,c_j^\dagger\} = \delta_{ij},\; \{c_i,c_j\}=\{c_i^\dagger,c_j^\dagger\} = 0 \, ,
\end{equation}
spin operators obey mixed commutation relations. On different sites, operators commute while on-site operators anti-commute. There exist different mapping methods between fermions and spin operators, such as the Jordan-Wigner transformation,\cite{JW28}
Majorana representation,\cite{Schad15} or the Bravyi-Kitaev transformation.\cite{BK02}
Here, we use the most common method, the one dimensional Jordan-Wigner transformation (JWT)\cite{JW28}
\begin{equation}\label{eq:JWTransformation}
 c_i = \prod_{j<i}\left(-\sigma_j^z\right)\sigma_i^- = e^{i\phi_{i}}\sigma_i^- \, .
\end{equation}
The JWT reproduces correct commutation relations and preserves the dimensionality of the Hilbert space.
This transformation requires ordering of the operators and, unfortunately, introduces non-local interactions between all the qubits of the quantum simulator. For example, the one-particle term $\sum_{ij}t_{ij}c_i^\dagger c_j$ transforms into
\begin{equation}
 H_q = \sum_{ij}\hat\Phi_{ij}t_{ij}\sigma_i^+\sigma_j^- \, ,
\end{equation}
 with the Jordan-Wigner string $\hat\Phi_{ij}=\prod_{k=\mathrm{min}(i,j)}^{\mathrm{max}(i,j)}(-\sigma_k^z)$ that accounts for the correct parity of the system.
The transformed Hamiltonian $H_q$ can can either be implemented on a digital quantum computer
or an analog quantum simulator\cite{MAN02}. Either way, from these simulations we can in principle extract the systems Green's functions, for example $G^<_{0,ij}(t,t')\propto\langle c_i^\pd(t) c_j^\dagger(t')\rangle_0$,\cite{Tian2016,Wilhelm2016} where the subscript 0 refers to the noise-free result.
Unfortunately, a realistic quantum simulator suffers from decoherence. Qubits couple to uncontrolled environmental degrees of freedom, such as two-level systems,\cite{MSS06} and information is lost to these environmental degrees of freedom.

Decoherence of single qubits and small systems of qubits has been analyzed with quantum master equations (QME). \cite{CAR02}
The QME approach allows for excellent quantitative results for small systems. Here, we want to qualitatively understand the behavior of large systems.
Instead of a master-equation approach we will map the faulty quantum simulator back to a fermionic system and expand the disturbed Green's function $G(\omega)$, obtained from the simulation on the faulty quantum simulator, in powers of the coupling to the noise. This is identical to an expansion in terms of the ideal Green's function of the unperturbed quantum simulator $G_0(\omega)$ which we would like to obtain from the simulation, and establishes the connection between perturbed and ideal Green's function, Eq.~\eqref{eq:Central_Result_in_Intro}.

\subsection{Decoherence as a fermionic problem}
The qubits of a quantum simulator couple to and thus dissipate information into the noisy environment.  
The interaction with the environment adds a perturbation of the form
\begin{equation}
 H_{q,\alpha} = \sum_i \sigma_i^\alpha\hat X_i^\alpha \, ,
\end{equation}
to the quantum simulator Hamiltonian, where $\alpha=x,\, z$
and $\hat X_i^\alpha$ is a bath operator that describes the coupling between qubit $i$ and its respective bath.
Longitudinal coupling $\propto \sigma^z_i$
describes (pure) dephasing, while transverse coupling $\propto\sigma_i^x$ is responsible for relaxation. 
Their dynamics are governed by free bath Hamiltonians $H_{B,\alpha}$. We assume that each qubit couples to individual, uncorrelated baths for decay and dephasing, such that
\begin{equation}\label{eq:baths}
	\langle \hat X_i^\alpha(t)\hat X_j^\beta(t')\rangle_0=\delta_{ij}\delta_{\alpha\beta}\langle \hat X_i^\alpha(t)\hat X_i^\alpha(t')\rangle_0
\end{equation}
holds for all free bath correlation functions.
Since Wick's theorem does not apply for spin operators, we map noise and quantum simulator Hamiltonians back to fermionic operators. For dephasing the Jordan-Wigner transformation, Eq.~(\ref{eq:JWTransformation}), simply yields an on-site term of the form
\begin{equation}\label{eq:dephasing_H}
	H_{f,z} = \sum_i c_i^\dagger c_i^\pd\hat X^z_i \, .
\end{equation}
This mapping is quite general and can be used for any noise inducing dephasing. 
While longitudinal coupling is easy to handle after the transformation, the transformation of transversal coupling induces non-local electron interactions of the form
\begin{equation}\label{eq:decay_H}
	H_{f,x} = \sum_i \prod_{j<i}(1-2c_j^\dagger c_j^\pd)(c_i^\dagger+c_i^\pd)\hat X_i^x  \, ,
\end{equation}
due to Jordan-Wigner strings. 
Additionally to non-locality, the mapping of transversal coupling of qubits to the bath with a Jordan-Wigner transformation produces a Hamiltonian that is odd in fermionic creation and annihilation operators. 
The linear term $\sim (c_i^\pd+c_i^\dagger)$ in the coupling stems from the violation of fermion number conservation due to the exchange of excitations between qubit and bath. 
This linearity induces another fundamental problem: One cannot replace the Dyson time-ordering that appears in the time evolution operator with a Wick time-ordering necessary for Wick's theorem.
This problem can be avoided if the bath operator $\hat X_i^x$ is odd in fermionic operators. In this case, the Hamiltonian becomes even in creation/ annihilation operators and we can use the standard perturbation theory with Wick time-ordering. In this paper, we discuss a system where this is the case. 
With dephasing, decay, and free bath Hamiltonian, the effective fermionic Hamiltonian that describes a noisy quantum simulator with fermionic operators takes the form
\begin{equation}\label{eq:H0}
	\mathcal H_f = H_f + H_{f,z}+H_{f,x}+H_{B}.
\end{equation} 
We use this Hamiltonian to describe a noisy quantum simulator with the Keldysh contour technique.

\section{Decoherence on Keldysh contour}\label{sec:KeldyshContour}
In this section, we calculate the non-equilibrium electronic Green's functions of the noisy quantum simulator. 
The general idea is to use a non-equilibrium field theory to expand the Green's functions in the coupling between system (quantum simulator) and bath to  obtain the Dyson equation shown in Eq.~\eqref{eq:Central_Result_in_Intro}. 
To check the validity of the mapping from a noisy quantum simulator to a many-body fermionic system, Eqs.~\eqref{eq:dephasing_H} and \eqref{eq:decay_H}, we compare the field theoretical results with master-equation calculations for small systems. 
We choose a non-interacting fermion system,
\begin{equation}
	H_f = \sum_{ij} t_{ij}c_i^\dagger c_j^\pd\,,
\end{equation}
in order to focus on effects of decoherence.
Due to the assumption given in  Eq. \eqref{eq:baths}, we can treat dephasing and decay separately. 
In the first part of the section, we briefly show how to calculate the Green's functions in a typical system-bath approach with the help of a master equation. 
On a classical computer, this method is feasible for small systems with a size of up to $~20-30$ qubits.
In the second part, we use the mapping of the noise to fermions and obtain the Green's functions in the presence of longitudinal coupling to a bath of harmonic oscillators with Ohmic spectral density.
In the last part of the section, we show how to calculate decay at zero temperature due to a bath of two level systems, one of the major sources of decoherence for superconducting 
qubits.\cite{MAR05,MUL15,BAR14}

\subsection{Master equation}
For small systems the time evolution of the noisy quantum simulator with Hamiltonian 
$\mathcal H_q=H_q+H_{q,x}+H_{q,z}+H_B$ 
can be simulated approximately with quantum master equations.
In the Lindblad form, the density matrix of the quantum simulator obeys the equation of motion
\begin{align}\notag
 \dot\rho = -&i[\rho,H_q] + \sum_n \frac{\Gamma_{2^*,n}}{2}(\sigma_n^z\rho\sigma_n^z-\rho)\\ \label{eq:qme}
 &+\frac{\Gamma_{1,n}}{2}(2\sigma_n^-\rho\sigma_n^+-[\sigma_n^+\sigma_n^-,\rho]_+) = \mathcal L \, \rho \, ,
\end{align}
with the dephasing rate $\Gamma_{2^*,n}=S_{z,n}(0)$, and the decay rate $\Gamma_{1,n}=S_{x,n}(\epsilon_n)$. The bath spectral density $S_n(\omega)$ will be defined below, in Eq.~\eqref{eq:spectral_density}.
We can formally integrate the master equation \eqref{eq:qme} and obtain $\rho(t)=e^{\mathcal L (t-t')}\rho(t')$, where $\mathcal L$ is the Lindblad super-operator.
From here we use the quantum regression theorem\cite{CAR02} to obtain two-time correlation functions. For $\tau\geq 0$ the theorem states
\begin{align}
 \langle \hat A(t+\tau)\hat B(t)\rangle &= \mathrm{tr}\{\hat A e^{\mathcal L\tau} [\hat B\rho(t)]\}\,,\\
 \langle \hat A(t)\hat B(t+\tau)\rangle &= \mathrm{tr}\{\hat B e^{\mathcal L\tau} [\rho(t)\hat A]\}.
\end{align}
We implement Eq.~\eqref{eq:qme} with the QuTip python package\cite{qutip} to calculate the disturbed fermionic Green's functions, for example,
\begin{align}\notag
 G&^>_{nn'}(\tau) = -i\langle c_n\pd(t+\tau)c_{n'}^\dagger(t)\rangle\\ 
 &=-i \left\langle \prod_{k<n}(-\sigma_k^z)\sigma_n^-(t+\tau)\prod_{l<n'}(-\sigma_l^z)\sigma_{n'}^-(t)\right\rangle\, .
\end{align}
Due to the presence of noise we expect differences to the ideal result.
The method can be applied straightforwardly to time dependent problems.

To compare to the calculations obtained in the steady-state approach of the fermionic theory (see below for details), we assume that in the distant past, at $t=0$, our system was decoupled from the bath and in thermal equilibrium $\rho(0)=Z^{-1}e^{-\beta H_q}$, where $\beta$ is the inverse temperature. 
Subsequently, we allow the system to evolve under the Hamiltonian $\mathcal H_q$ for a time $t\gg \Gamma_0^{-1}$, with $\Gamma_0$ the smallest decoherence rate. 
During this evolution correlations between bath and qubit system build up and the entire system reaches a steady state at time $t$ with density matrix $\rho_0\equiv\rho(t)$. 
Because the system has reached a stationary state at time $t$, the Green's function depends only on the time difference $\tau$, while the exact starting time $t\gg \Gamma_{2^*}^{-1}$ is insignificant. 
Thus, we can calculate the Fourier transform with respect to $\tau$ and compare the results with the Green's function obtained with the fermionic field theory.

\subsection{Fermionic perturbation theory}

The results of a noisy quantum simulation are encoded in the perturbed non-equilibrium Green's function (matrix in the Keldysh space)
\begin{equation}\label{eq:Greens_Definition}
	G(t,t') = \begin{pmatrix}G^+(t,t')&G^K(t,t')\\0&G^-(t,t')\end{pmatrix}\,,
\end{equation}
where $G^\pm$ are the retarded (+) and advanced (-), and $G^K$ the Keldysh Green's function, defined as
\begin{align}
	G_{ij}^\pm(t,t') &= \mp i\theta(\pm(t-t'))\left\langle \left[c_i(t),c_j^\dagger(t')\right]_+\right\rangle\,,\\
	G_{ij}^K&=-i\left\langle \left[c_i(t),c_j^\dagger(t')\right]_-\right\rangle\, .
\end{align}
Due to the mapping between quantum simulator and fermionic system the fermionic Green's function corresponds to certain qubit correlators that can be measured in an experiment. For example,
\begin{equation}
	G_{ii}^K(t,t') = -i\langle (\sigma^-_i(t)\sigma^+_i(t')+\sigma^+_i(t')\sigma^-_i(t)\rangle \, .
\end{equation}
To obtain Eq.~\eqref{eq:Central_Result_in_Intro}, which connects the set of perturbed
Green's functions, Eq.~\eqref{eq:Greens_Definition}, to the ideal ones, we follow Refs.~[\onlinecite{RS86},\onlinecite{SvL13}] and first introduce the contour-ordered Green's function
\begin{align}\notag
 iG^C_{ij}(t,t') &= \left\langle T_C\, c_i(t)c_j^\dagger(t')\right\rangle\\\label{eq:time_ordered_G}
 &= \Tr\left[ T_C\, U_C\, \tilde c_i(t)\tilde c_j^\dagger(t')\rho(z_0)\right],
\end{align}
which can be reduced to the real-time Green's functions Eq.~\eqref{eq:Greens_Definition}.
Here, $T_C$ is the time-ordering operator along the contour
and $\rho(z_0)$ is the density matrix of quantum simulator and environment at the initial contour time $z_0$.
In what follows we require that the density matrix at $z_0$ factorizes into bath and system $\rho(z_0)=\rho_S(z_0)\otimes\rho_B(z_0)$ and the system is non-interacting, e.g.,
\begin{equation}
 \rho_S(z_0) =Z_S^{-1} e^{-\beta \sum_{ij}a_{ij}c_i^\dagger c_j^\pd}=Z_S^{-1} e^{-\beta H_{\rm ini}} \,,
\end{equation}
while the bath is in thermal equilibrium.
This condition ensures the validity of Wick's theorem, which we will use in the next step.
The exact choice of the contour depends on initial conditions of the problem. The most general contour is depicted in Fig.~\ref{fig:SE}(a).
We comment on the different choices for the contour in the following subsections. 
To obtain the second equality in Eq.~\eqref{eq:time_ordered_G} we changed to the interaction picture with respect to the coupling between simulator (system) and environment (bath) where operators in the interaction picture are defined as
\begin{equation}
 \tilde O(t) = e^{i (H_f+H_B) t}Oe^{-i (H_f+H_B) t} \, .
\end{equation}
We further introduced the time evolution along the contour in the interaction picture,
\begin{equation}\label{eq:contour_evolution}
 U_C = T_C\exp\left\{-i\int_C dt'[\tilde H_{f,x}(t')+\tilde H_{f,z}(t')]\right\} \, ,
\end{equation}
where the time integration runs along the contour $C$. 
We expand $U_C$ in powers of the interaction between bath and system, apply Wick's theorem to both bath and system operators.
We obtain a diagrammatic expansion of the full contour-ordered Green's function in terms of the ideal contour-ordered Green's function $G_{C.0}$ and bath correlation functions $D_C$.
Defining the self-energy $\Sigma_C$ as the sum of all irreducible diagrams we find the contour Dyson equation
\begin{align}\notag
	\hat G_C(z_1,&z_2) = \hat G_{C,0}(z_1,z_2) \\ \label{eq:Dyson_contour}
	&+\int_C d\bar zd\bar z' \,\hat G_{C,0}(z_1,\bar z)\Sigma(\bar z,\bar z')\hat G_C(\bar z',z_2) \, .
\end{align}
An analytical continuation from contour to real times yields Dyson equations for the non-equilibrium Green's functions of the noisy quantum simulator introduced in Eq.~\eqref{eq:Greens_Definition}.

\subsubsection{Interacting initial state}
The full contour depicted in Fig.~\ref{fig:SE}(a) covers the most general situation:
At the start of the simulation, $z=t_0$, we prepare the system in an initial density matrix $\rho(t_0)$ of the simulator. 
Because the initial density matrix on the contour has to be non-interacting, we have to include the time 
evolution along the vertical path from $z_0=t_0-i\beta$ with some initialization Hamiltonian $H_{M}(z)$ that evolves the non-interacting density matrix $\rho(z_0)$ into the interacting matrix at $t_0$.
Here, $H_M$ is a Hamiltonian defined in such a way that it evolves the system to the correct initial state along the vertical branch.\cite{SvL13}
We will not use this contour because we find that we can treat many experimentally relevant situations with simpler contours.

\subsubsection{Non-interacting initial state}
We assume that we prepare the simulator in an initial state that can be cast into the non-interacting form
\begin{equation}
	\rho(t_0) = e^{-\beta\sum_{ij}a_{ij}c_i^\dagger c_j}=e^{-\beta H_\mathrm{ini}}.
\end{equation}
With this kind of initial state we can cover a lot of interesting situations.\cite{SvL13} For example with $a_{ij} = \delta_{ij}(1-2\delta_{i1})$ and $\beta\to\infty$ we can implement a situation where qubit $i=1$ is in the up state while all remaining qubits are in the down state.
Without initial correlations and non-interacting initial density matrices of the form introduced above, we can omit the vertical part of the contour (a) and evolve only along the real time forward and backward branch of the contour.
For this case, we can cast the contour Dyson equation \eqref{eq:Dyson_contour} into differential equations for the real-time Green's functions \eqref{eq:Greens_Definition}:
\begin{align}\label{eq:KDE_1}
	\left(i\frac{d}{d t}-\hat H_f(t)\right)G^{\pm}(t,t')
	 &= [\Sigma^\pm\circ G^\pm](t,t')\,,\\
	 \left(i\frac{d}{d t}-\hat H_f(t)\right)G^{K}(t,t')
	 &= I^K(t,t')\,,\\
	I^K(t,t') = [\Sigma^+\circ G^K&+\Sigma^K\circ G^-](t,t') \, ,
\end{align}
where we introduced the convolution $[A\circ B](t,t') = \int_{t_0}^T d\bar t\, A(t,\bar t)B(\bar t, t')$. Due to the structure of the contour, it suffices to set the upper limit of the convolution integrals to $T = \rm max(t,t')$.
This set of differential equations has to be solved with boundary conditions
\begin{align}
	G^K(0,0) = - i[1-2f(\hat H_{\rm ini})] \, ,
\end{align}
where $f(\hat H_{\rm ini}) = [1+\exp(\beta \hat H_{\rm ini})]^{-1}$.
Together with the identities
\begin{align}\label{eq:hermitianconjugate}
	G^+(t,t')&=\left[G^-(t',t)\right]^\dagger\\
	\label{eq:KDE_2}
	G^K(t,t')&=-\left[G^K(t',t)\right]^\dagger \, ,
\end{align}
we find a closed system of differential equations which can be solved either numerically or analytically.
These equations are known as the Kadanoff-Baym equations (KBE).
Except for special situations, the Green's functions will depend on the initial time $t_0$ and on both time arguments $t$ and $t'$ separately.
This approach corresponds to time-dependent simulations on an analog quantum simulator.

\subsubsection{Adiabatic switching}
\begin{figure}
\centering
\includegraphics[]{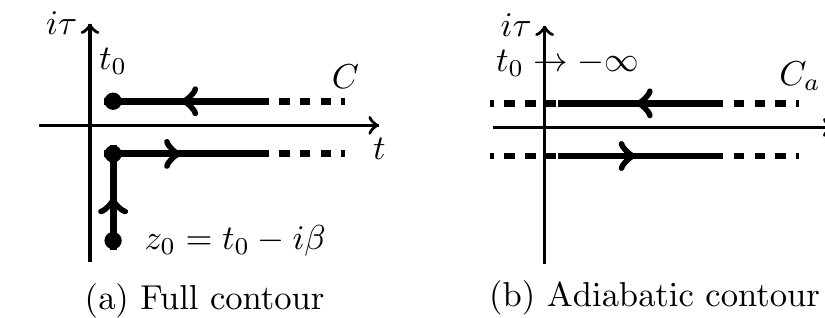}
\includegraphics[]{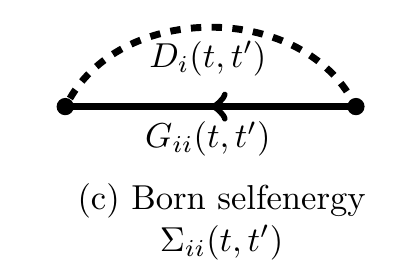}
\caption{\label{fig:SE}(a) full contour $C$ and (b) adiabatic contour $C_a$. (c) Born self-energy $\Sigma_{ii}$ for longitudinal coupling to harmonic oscillators.}
\end{figure}

If we are not interested in time-dependent simulations but only on equilibrium properties such as the spectral function or decoherence rates, we can further simplify the contour.
Since we are not interested in the evolution from some initial time $t_0$ we let $t_0\to-\infty$ and evolve the system on the contour \ref{fig:SE}(b).\cite{KAM11}
In the distant past, system and bath were decoupled and in thermal equilibrium at temperature $T$.
We switch the coupling between the bath and the system adiabatically from zero to the full value, i.e. $(H_{f,x}+H_{f,z}) \to (H_{f,x}+H_{f,z})e^{-\eta |t|}$.
For a time-independent Hamiltonian $H_f$ this approach describes a steady state of the system. We assume that the Green's function depends on the time difference only $G(t,t') = G(t-t')$, while components that depend on the center of mass time $\sim t+t'\gg t-t'$ have decayed.
This corresponds to neglecting initial correlations of the system.
In this case we can introduce the Fourier transformed function $G(\omega) = \int d t e^{-i\omega t}G(t)$.
The Dyson equation for the Fourier transformed Green's functions simplifies to Eq. \eqref{eq:Central_Result_in_Intro}:\cite{KAM11}
\begin{align}
 G^{\pm}(\omega)&=\left[(G_0^{\pm})^{-1}(\omega)-\Sigma^\pm(\omega)\right]^{-1}\\
 G^K(\omega) &= -G^+(\omega)\Sigma^K(\omega)G^-(\omega) \, .
\end{align}
From retarded and advanced Green's function follows the spectral weight 
\begin{align}
 \mathcal A(\omega)=i[G^+(\omega)-G^-(\omega)],
\end{align}
a quantity we would like to extract in an experimental simulation.
We use relation~\eqref{eq:hermitianconjugate} between retarded and advanced self-energy and express $\Sigma^{\pm}= \hat\Delta(\omega)\pm i\Gamma(\omega)/2$ through the rate function $\hat\Gamma(\omega)=2\Im[\Sigma^+(\omega)]$ and the energy shift $\hat\Delta(\omega)=\Re[\Sigma^+(\omega)]$.
The rate corresponds to the lifetime of excitations and determines the width of peaks in the spectral function.
It compares to decoherence rates used in master equation calculations.
We note that for a time independent Hamiltonian $\hat H_f$ the retarded and advanced Green's function always depend only on the time difference $G^\pm(t,t')=G^\pm(t-t')$. This statement is independent from initial state preparation.
Thus, for any time independent Hamiltonian we can define the Fourier transformed retarded and advanced functions.
The Keldysh component $G^K$ carries information about the occupation of states and is connected to the distribution function $F$ of the system as \cite{KAM11}
\begin{align}
 G^K(t,t') = G^+\circ F(t,t')-F\circ G^-(t,t').
\end{align}
Thus, from $G^K(t,t)$ follows the distribution $F(t)$ at a given time $t$.
We conclude this section by introducing the ideal Green's functions of the fermionic system
\begin{align}
 G_0^\pm(\omega) &= \left[\omega -\hat H_f \pm i0\right]^{-1}\\
 G_0^K(\omega) &= [G^+_0(\omega)-G_0^-(\omega)]F(\omega)
\end{align}
where $F(\omega)=1-2f(\omega)$ is the fermion distribution function, $f(\omega)=[1+\exp(\beta \omega)]^{-1}$ is the Fermi function
and $\hat H_f$ is the matrix representation of the free Hamiltonian $H_f$ in the local qubit basis.

\section{Adiabatic approach and steady-state analysis}\label{sec:SteadyState}
In this section, we analyze systems in the adiabatic limit. We focus on decoherence and thus consider a non-interacting fermionic system.
We calculate dephasing rates due to bosonic environment, 
analyze relaxation due to a bath of two-level systems, and demonstrate the influence of single qubit decoherence on the spectral density of large simulated systems.
For small systems we compare the results obtained with our method with master-equation calculations.

\subsection{Dephasing due to a bosonic bath}
Here, we calculate the Green's functions for a quantum simulator subject to pure dephasing due to a bath of harmonic oscillators,
where the qubits couple linearly to the displacement of the oscillators
\begin{equation}\label{eq:dephasing_oscillators}
	\hat X_i^z = \sum_s g_{is} (a_{is}^\pd+a_{is}^\dagger) \, .
\end{equation}
This model describes, e.g., coupling of a qubit to a resistive environment.\cite{CL87} 
The bath of oscillators is characterized by its power spectral density\cite{MSS05}
\begin{equation}\label{eq:spectral_density}
	S_{i}(\omega) = \left\langle \left\{\hat X_i^z(t),\hat X_i^z(t')\right\}\right\rangle_{0,\omega}= J_i(\omega)\coth\frac{\beta\omega}{2} \, ,
\end{equation}
where $J_i(\omega)$ is the spectral function and $\langle\cdots\rangle_0$ is an average with respect to $\rho_B(t_0\to-\infty)$.

The interaction term in Eq.~\eqref{eq:dephasing_H} with the noise operators in Eq.~\eqref{eq:dephasing_oscillators} is identical to a local interaction between electrons and phonons. For this type of system the lowest-order self-energy in Born approximation for the interaction with the bath [see Fig.~\ref{fig:SE}(b)] 
is given by\cite{RS86}
\begin{align}
 i\hat\Sigma^{\pm}_{ij}(t,t') &= \delta_{ij}[iG^{\pm}_{ii}iD^k_i(t,t')+iG^{K}_{ii}iD^\pm_i(t,t')]\\ \notag
 i\hat\Sigma^k_{ij}(t,t')&=\delta_{ij}\{[G_{0,ii}^+(t,t')-G_{0,ii}^-(t,t')]\\
 &\hspace{-4em}\times[D_{i}^+(t,t')-D_{i}^-(t,t')]+iG^k_{ii}(t,t')iD^k_i(t,t')\}
\end{align}
Because bath operators of different qubits are uncorrelated, the self-energy is diagonal in the local qubit basis $\Sigma_{ij}\sim\delta_{ij}\Sigma_{ii}$. 
With the bath correlation functions
\begin{align}
	D_i^k (\omega) &= -iJ_i(\omega)\coth\frac{\beta\omega}{2}\\
	D^{\pm}_i(\omega) &= \int\frac{d\nu}{2\pi}\frac{J_i(\nu)}{\omega-\nu\pm i0} \, .
\end{align}
we find the diagonal components of the self-energy
\begin{align}\notag
	\Sigma_{ii}^{\pm} (\omega) = &-\frac 12 \int\frac{d\nu}{2\pi}\Bigl[G_{0,ii}^{\pm}(\omega-\nu)J_i(\nu)\coth\frac{\nu}{2T}
	\\\label{eq:sigma_full} 
	&+i\int \frac{d\nu'}{2\pi}G_{0,ii}^K(\omega-\nu)\frac{J_i(\nu')}{\nu-\nu'\pm i0}\Bigr]\,,\\ \notag
	\Sigma^K_{ii}(\omega) = &-\frac 12 \int\frac{d\nu}{2\pi}J_i(\nu)(G_{0,ii}^+(\omega-\nu)-G_{0,ii}^-(\omega-\nu))\\
      &\times\Bigl[F(\omega-\nu)\coth\frac{\nu}{2T}+1\Bigr].
\end{align}
From the self-energy we extract the lifetime
\begin{align}\notag
 \hat\Gamma_{ii}(\omega)=&-\frac 12 \int\frac{d\nu}{2\pi}\mathcal A_{0,ii}(\omega-\nu)) J_i(\nu)\\&\times\left[F(\omega-\nu)+\coth\frac{\nu}{2T}\right]
\end{align}
with the spectral weight $\mathcal A_0(\omega)$ of the unperturbed simulator.
In the following we calculate the self-energy and rate function for a single qubit.
For a flat spectral density we expect to find the golden rule dephasing rate $\Gamma_2 \propto S(0)$. 
After this we analyze a system of coupled qubits and compare the results with master-equation calculations.

\subsubsection{Single Qubit Dephasing}
The dephasing rates due to noise with a flat spectral density can be calculated with Fermi's golden rule.
Within the same approximations the rates calculated with the non-equilibrium Green's functions must certainly yield the same dephasing rates.
For a single qubit with energy splitting $\epsilon$ the free Green's functions are $G_0^\pm=(\omega-\epsilon\pm i0)^{-1}$ and $G^K_0=-2\pi i F(\omega) \delta(\epsilon-\omega)$.
The rate function takes then the form
\begin{align}
 \hat\Gamma(\omega)=\frac12J(\omega-\epsilon)[F(\epsilon)+\coth\frac{\beta(\omega-\epsilon)}{2}].
\end{align}

The Green's function for a single qubit is strongly peaked for energies close to the qubit energy splitting. 
As long as the spectral density $S(\omega)$ is not strongly peaked around $\omega=0$ and we have no static component,
i.e.~$J(0)\to0$, we can approximate the self-energy with its value at $\omega\approx \epsilon$.
This yields the single-qubit dephasing rate
\begin{align}\label{eq:rates_dephasing}
 \Gamma_2\approx \hat\Gamma(\epsilon)=\frac{S(0)}{2}\,.
\end{align}
This result corresponds to the golden-rule single-qubit dephasing rate due
to a bath characterized by a flat spectral density.

\subsubsection{Coupled System}
Now, we calculate the self-energy for coupled but non-interacting electrons with $H_f = \sum_{ij}t_{ij}c_i^\dagger c_j^\pd$.
We can diagonalize the Hamiltonian with a transformation
\begin{equation}
	c_i \to U_{ik}a_k
\end{equation}
where $U$ is a unitary transformation matrix and the Hamiltonian is diagonal in $a_k$, $H_f = \sum_k \epsilon_k a_k^\dagger a_k\pd$. 
We express the Green's functions of the fermions in the original basis with the Green's functions $\tilde G_{kk'}$ of the diagonalized system as
\begin{equation}
	G^\alpha_{ij} = \sum_{kk'}U_{k'j}^\dagger U_{ik}\tilde G^a_{kk'}(t,t').
\end{equation}
For the on-site ideal Green's functions this relation simplifies to
\begin{equation}
	G^a_{0,ii} = \sum_{k}|U_{ik}|^2\tilde G^a_{0,k}(t,t')\,,
\end{equation}
where the Green's functions in the diagonal basis take the same form as the single-qubit Green's functions with $\epsilon\to\epsilon_k$. 
We find the rate function
\begin{align}\label{eq:dephasing_sigma}
 \hat\Gamma_{ii}(\omega)=\frac12\sum_{k}|U_{ik}|^2 J_i(\omega-\epsilon_k)[F(\epsilon_k)+\coth\frac{\beta(\omega-\epsilon_k)}{2}]
\end{align}
For a flat spectral density, for which $J(\omega) \coth\frac{\omega}{2T}\approx \mathrm{const.}$,
the rate is frequency independent and we expect the Lindblad master equation to yield identical results as the Keldysh calculations.
If the spectral density varies on the frequency range determined by the rate function, the rate function will be frequency dependent.
Consequently, the constant-rate master equation does not yield the same results anymore.
In this situation,
we use a Bloch-Redfield master equation which evaluates the spectral density at the systems energies instead of the simple Lindblad equation.

\begin{figure}[tb]
	\includegraphics[width=\columnwidth]{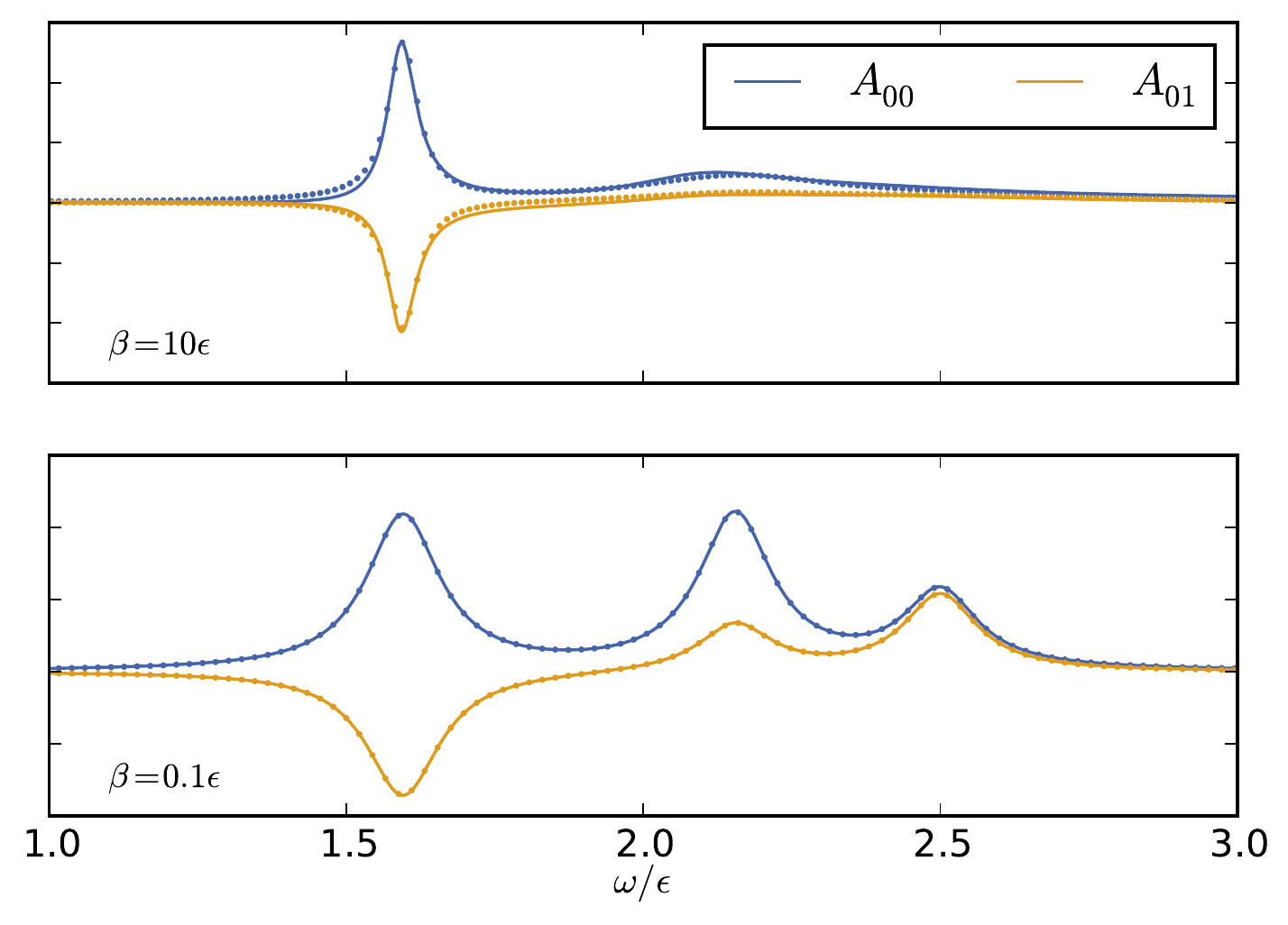}	\caption{\label{fig:stationary_dephasing}Numerical simulation of a tight binding chain with $N=5$ sites coupled to ohmic baths.
	 We plot the spectral weight $\mathcal A_{00}$ and $\mathcal A_{01}$ obtained via Keldysh calculation (solid line) and QME calculations (dots) for two different temperatures of the bath $T=\beta^{-1}$. For the upper plot we used a Bloch-Redfield master equation while the lower plot is obtained with a Lindblad master equation.}
\end{figure}
To compare the results of the fermionic theory with master-equation calculations, we choose a simple system: a linear chain with on-site energies $\epsilon$ and nearest neighbor hopping $t_{i,i+1}=g$,
\begin{equation}
 H_f = \sum_i \epsilon\, c_i^\dagger c_i^{\phantom\dagger}+\frac g2\, (c_{i+1}^\dagger c_i^{\phantom\dagger}+c_{i}^\dagger c_{i+1}^{\phantom\dagger}) \, ,
\end{equation}
with eigenenergies $\epsilon_k = \epsilon + g\cos k$, $k=2\pi n/N$.
For an ohmic bath, the condition of a flat spectral-density is fulfilled for high temperatures $T\gg \epsilon+g$.
In Fig.~\ref{fig:stationary_dephasing}, we compare master-equation calculations with the fermionic Green's functions obtained from Eq.~\eqref{eq:Central_Result_in_Intro}. 
In case of the high temperature system we use a Lindblad master equation with the single qubit dephasing rates $\Gamma_2$
from Eq.~\eqref{eq:rates_dephasing}.
The fermionic self-energy is given by Eq. \eqref{eq:dephasing_sigma}. We compare results for two different bath temperatures $T\gg \epsilon + g$ and $T\ll \epsilon + g$.
As expected, the Lindblad master equation and Keldysh theory are identical for high temperatures while the description with single qubit dephasing rates fails for a low temperature bath, i.e.~frequency dependent self-energy. 
Thus, for the low temperature situation (upper plot) we used QuTip's Bloch-Redfield implementation to obtain the master equation results.
Deviations between Bloch-Redfield and Keldysh results are due to the fact, that the spectral density vanishes very fast for $\omega\approx0$ and the
Bloch-Redfield solutions become unreliable.
We conclude, that the mapping of dephasing noise to fermionic operators works as expected and compares well to usual master-equation calculations.

\subsubsection{Effect of dephasing on large systems}
\begin{figure}
	\includegraphics[width=\columnwidth]{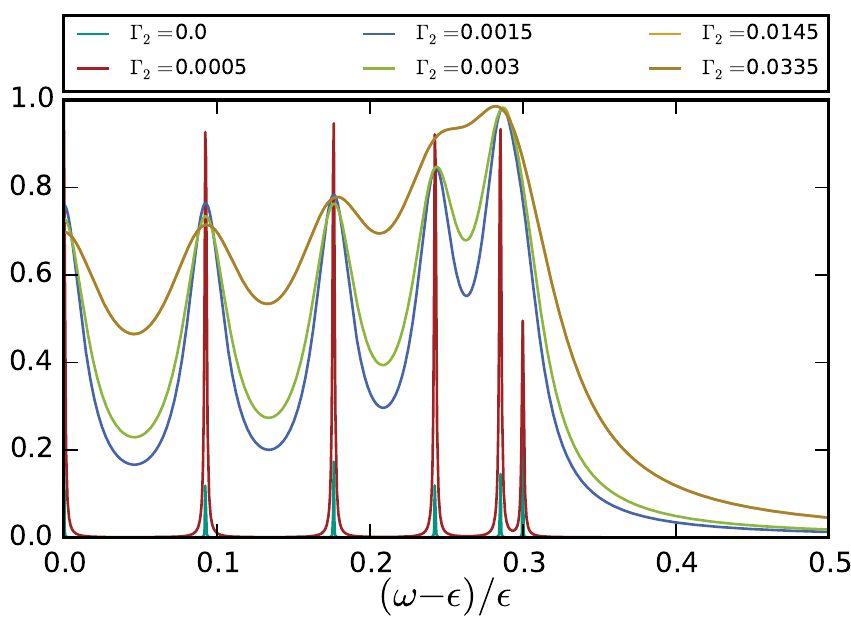}
	\vspace{-0.2cm}
	\includegraphics[width=\columnwidth]{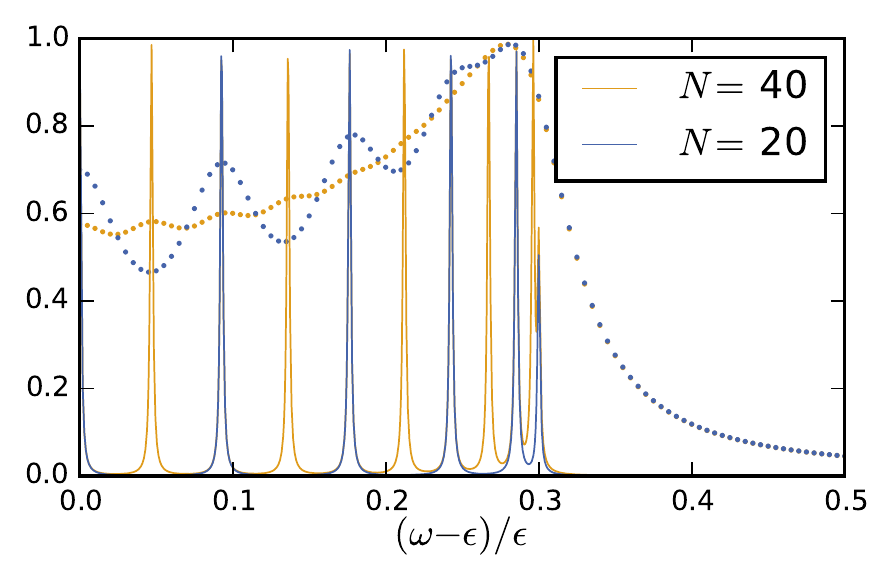}
	\caption{\label{fig:stationary_many}Numerical simulation of tight binding chains coupled to high temperature ohmic baths. (top)
  We plot the spectral weight $A_{00}=-i\pi \Im[G_{00}^R]$ for $N=20$ sites and increasing dephasing rates $\Gamma_2$.
  (bottom) For $N=20$ and $40$ sites we compare the spectral weight returned from a noisy quantum simulator (dots) with the ideal one (solid lines).}
\end{figure}

Another interesting question is the spectral resolution of a noisy quantum simulator. 
We expect intuitively that the resolution is determined by the decoherence rates $\Gamma_2$ of the single qubits.
For larger systems the eigenenergies become more and more dense and we expect an increasing loss of information due to decoherence.
To check the validity of these assumptions we calculate the retarded component of the perturbed Green's function for different dephasing rates $\Gamma_2$ and increasing system size.
In Fig.~\ref{fig:stationary_many}, we plot the spectral weight of a tight-binding chain for increasing single-qubit dephasing rates in the top graph. 
As expected, the perturbed quantum simulator can only resolve features further apart than the dephasing rates and the resolution decreases with increasing dephasing rate.
In the bottom graph, we show the influence of system size on the loss of information by comparing the perturbed spectral function (dots) with the ideal one (solid) for different system sizes. Here, we also confirm the expectation that
while the disturbed quantum simulator can clearly resolve most of the peaks for the smaller system ($N=20$), more peaks vanish for a larger system ($N=40$).

\subsection{Relaxation}\label{sec:stationary_decay}
Here, we show how the mapping to fermions works when describing  decay for a specific type of noise. 
Namely, we use the Jordan-Wigner transformation to calculate relaxation due to a bath of two level systems at low temperature. 
Although academic at first glance, this describes a relevant problem. 
Experimentally and theoretically it is well understood, that two-level systems (TLS) are responsible for a large part of decoherence in superconducting devices.\cite{MAR05} 
For example, the omnipresent $1/f$ noise can be attributed to a bath of TLS with homogeneously distributed properties.\cite{MSS06}
TLS are low energy excitations such as tunneling atoms, dangling bonds or trapped electrons inside amorphous and dielectric layers of superconducting devices.\cite{COL10,LIS16}
The TLS couple via their dipole moment to the electrical field inside the qubit junction and the coupling takes the form
\begin{equation}\label{eq:HcDecay}
	\hat X_i^x=\sum_s g_{is}(\sigma^+_i\tau_{is}^-+ {\rm h.c.}) \, ,
\end{equation}
where $\sigma^{\pm}$ are qubit operators and $\tau^{\pm}$ are pauli matrices in TLS space. 
Mapping only qubit operators to fermions, we explicitely violate fermion number conservation, see Eq.~\eqref{eq:decay_H}.
For this particular model we can circumvent this problem by mapping both sets of spin-$\frac12$ operators, i.e. qubits and TLS, to fermions according to Eq. \eqref{eq:decay_H}.
As we will show below, this approach ensures particle conservation and we can use standard non-equilibrium perturbation theory.
In order to use a Jordan-Wigner transformation on both types of Pauli operators, we have to order the TLS in some way together with the qubits. 
We choose the following convention: We label the first qubit as fermion number one, $\sigma_1^-\to c_1$. 
Then follow all $N_1$ TLS that couple to the first qubit according to $\tau^-_{1,s}\to a_{1,s}$. The next fermion is the second qubit $c_2$ at overall fermion position $1+N_1+1$ and so on. With this numbering, the JWT of the different parts of the Hamiltonian reads
\begin{align}
 \sigma_i^-&\to\phi_i c_i\\
 \tau_{i,s}^-&\to \phi_n\varphi_{s-1}^{(i)}a_{i,s}\\
 \varphi_s^{(i)}&=\prod_{\alpha=1}^{s-1}(1-2a_{i,s}^\dagger a_{i,s}^{\phantom\dagger})\\
 \phi_{i+1} &= \phi_i(1-2c_i^\dagger c_i^{\phantom\dagger})\varphi_{N_i}^{(i)},\quad\phi_1 = 1 \, ,
\end{align}
where $N_i$ is the number of TLS coupling to qubit $i$.
The JWT introduces many-particle interactions between TLS fermions and qubit fermions. 
Both qubit-qubit hopping as well as interactions between qubits and TLS are affected by this interaction and the entire problem becomes intractable.
However, TLS responsible for decay have energies $\epsilon_{\rm TLS}$ close to typical qubit energies and are thus at low temperatures in typical experimental situations. For TLS at low temperatures $T_B\ll \epsilon_{\rm TLS}$ we can simplify the transformed Hamiltonian because the expectation values for the TLS occupation number $\hat n_{is}$ is negligible small and
we can safely let $\hat n_{is}\to0$ in the Hamiltonian. With this we recover the original qubit Hamiltonian with an additional effective interaction of the form
\begin{equation}
	H_{f,x}\approx \sum_{i,s}g_{is}c_{i}^\dagger a_{is}^{\phantom\dagger}+ {\rm H.c}
	\equiv {\bf c}^\dagger \hat T {\bf a} + {\rm h.c.} \, .
\end{equation}
This Hamiltonian describes tunneling of particles between two fermionic leads, one comprised of the TLS fermions $a_{is}$ and one of qubit fermions $c_i$ with tunneling amplitude $g_{is}$. To obtain the last equality we introduced vectors ${\bf c} = (c_1,\dots,c_N)^{\rm T}$ and a tunneling matrix $\hat T$.

Because the Hamiltonian is quadratic in fermionic operators we can exactly calculate the Green's functions.
Extending the tunneling matrix to Keldysh space we find that the Keldysh component of the tunneling matrix $\hat T$ vanishes, i.e.~$T^K = 0$.
Expanding the qubits Green's function in the tunneling matrix we find
\begin{align}\notag
	G &= G_0+ G_0 \hat T G_B \hat T^\dagger G_0 + G_0 \hat T G_B \hat T^\dagger G_0\hat T G_B \hat T^\dagger G_0+\cdots   \\ 
	 &= G_0+G_0\Sigma G   \, ,
\end{align}
with self-energy $\Sigma = \hat TG_B\hat T^\dagger$, where $G_B$ is the free TLS Green's function.
Here, $G$, $G_B$, $\hat T$, and $\Sigma$ are matrices in Keldysh space.
This type of self-energy is known as an embedding self-energy in the context of small systems such as quantum dots, coupled to large fermionic systems, e.g., electronic leads.\cite{SvL13} 
Using that the tunneling matrix is diagonal in qubit space we find the Dyson equation
\begin{align}
	(G^{-1})^\pm_{ij} &= (G_0^{-1})^\pm_{ij}-\Sigma^\pm_{ij} \\ \label{eq:sigma_decay}
	\Sigma^\pm_{ij} &= \delta_{ij}\sum_s |g_{is}|^2G^\pm_{B,is} \, ,
\end{align} 
where $a=\pm,K$ denotes the components in Keldysh space.
Defining the TLS spectral density $J_i(\omega)=2\pi\sum_s |g_{is}|^2\delta(\omega-\omega_{is})$ we find the retarded and advanced Green's functions
\begin{align}\notag
	[(G^{\pm})^{-1}]_{ij}
	&=[(G^{\pm}_0)^{-1}]_{ij}-\delta_{ij}\int_0^\infty \frac{d\omega'}{2\pi}\frac{J_i(\omega')}{\omega-\omega'\pm i0}\\
	&\approx [(G^{\pm}_0)^{-1}]_{ij}\pm i\delta_{ij}\frac12J_i(\omega)\,.
\end{align}
In the last step, we have neglected the real part of the self-energy.
We identify the decay rate $\hat\Gamma_i(\omega)=J_i(\omega)$.
To calculate the Keldysh component of the Green's function we employ the fluctuation-dissipation relation
\begin{equation}
   \Sigma^K=[\Sigma^+-\Sigma^-]F_{\rm tls} = -i\hat\Gamma(\omega)F_{\rm tls}\,,
\end{equation}
with the TLS distribution function $F_{\rm tls}(\omega)=\tanh(\beta_B\omega/2)\approx1$.
We calculate the Keldysh component of the Green's function with the relation
\begin{equation}\label{eq:decay_keldysh}
	G^K = -G^+\Sigma^KG^-\,.
\end{equation}
For uncoupled qubits, e.g. $t_{ij}=\epsilon_i\delta_{ij}$ we find
\begin{align}\label{eq:gr_decay}
	G^{\pm}_i(\omega) &\approx (\omega-\epsilon\pm i\Gamma_i)^{-1}\\ \label{eq:gk_decay}
	G^K_i(\omega) &= [G^+(\omega)-G^-(\omega)]\tanh\left(\frac{\omega}{2T}\right)\,,
\end{align}
with $\Gamma_i = J_i(\epsilon_n)$. The rates $\Gamma_i$ correspond to Fermi's golden-rule decay-rates for single qubits coupled to a bath of TLS.
Looking at Eq.~\eqref{eq:gk_decay}, we see that the qubit system equilibrates to the bath temperature $T\approx 0$ and obtains a finite lifetime due to interactions with the bath.

\section{Relaxation - Transient behavior}\label{sec:time_dependence}
In the previous section, we analyzed stationary-state properties of a quantum simulator coupled to the environment. The stationary state analysis yields information such as decoherence rates of the system.
However, many experiments on quantum simulators are far away from stationary states.
In a typical experiment, we prepare the quantum simulator in a certain initial state that is not necessarily a stationary state of the quantum-simulator Hamiltonian.
For a general initial state $\rho_0 = \ket\psi\bra\psi$ the Green's functions depend on the initial state configuration and on both time arguments independently.
We use the full contour without the vertical branch according to the Kadanoff-Baym equations (KBE), Eqs.~(\ref{eq:KDE_1}-\ref{eq:KDE_2}).

For the model discussed in section \ref{sec:stationary_decay}, a quantum simulator coupled to a bath of non-interacting TLS at zero temperature, 
we have already derived an expression for the self-energy, Eq. \eqref{eq:sigma_decay},
\begin{equation}
	\Sigma^\pm_{ij} = \delta_{ij}\sum_\alpha |g_{is}|^2G^\pm_{B,i\alpha} \, .
\end{equation}
Since the TLS are in thermal equilibrium at, $T_B=0$, we can use the fluctuation-dissipation relation to express the Keldysh component of the self-energy as
\begin{equation}
	\Sigma^K = \Sigma^+-\Sigma^-.
\end{equation}
Numerical integration of the KBE with this self-energy is straightforward.
Assuming that the relevant energies of the quantum simulator lie well within the excitation spectrum of the TLS, we can use the wide band approximation for the TLS spectral density, $J_i(\omega) =\sum_s |g_{is}|^2\delta(\omega-\epsilon_{is})=\Gamma_i = \mathrm{const.}$. In this approximation the bath of TLS corresponds to a white noise with time-local self-energies
\begin{align}
	\Sigma^{\pm}_{ii}(t,t') &\approx \mp i \frac{\Gamma_i}{2}\delta(t-t')\,,\\
	\Sigma^K_{ii}(t,t') &\approx i\Gamma_i\delta(t-t') \, ,
\end{align}
where the rate $\Gamma_i$ coincides with the decay rate $\Gamma_{1,i}$ used in a master-equation approach.
This approximation for the self-energy, known as the Markov approximation, is valid as long as internal TLS decoherence times are fast compared to typical system time scales.
Thus, we can easily compare the results obtained via KBE with master equation calculations.
For a time-independent Hamiltonian $\hat H_f$ retarded and advanced components depend only on the time difference and we can define the spectral weight $\mathcal A(\omega)=i(G^+(\omega)-G^-(\omega))$.
The retarded and advanced Green's function are identical to the equilibrium functions \eqref{eq:gr_decay}.
For time-independent Hamiltonian we find the equal time kinetic function
\begin{align}\notag
 &G^K(t,t) = -i\int\frac{d\omega}{2\pi}\bigl\{\mathcal A(\omega)F_{\rm tls}(\omega)-e^{-i\hat H_f t-\hat\Gamma/2t}\\
 &\times\left[\mathcal A(\omega)F_{\rm tls}(\omega)-\delta(\omega-H_{\rm ini})F(\omega)\right]e^{i\hat H_f t-\hat\Gamma/2t}\bigr\}\,,
\end{align}
which is directly related to the (equal time) time dependent distribution function $F(t,t)$ of the simulator.
Here, $[\hat\Gamma]_{ij}=\delta_{ij}\Gamma_i$ is the matrix of decay rates.
This result is quite intuitive. For $t=0$ the qubit system is described by the initial distribution $F(\hat H_{\rm ini})$ specified by the preparation of the simulator.
Subsequently, the simulator evolves with $\hat H_f$ under the influence of the environment.
The environment exchanges excitations with the simulator leading to a decay of the initial distribution to a stationary value.
For long times $t\gg \text{min}(\hat\Gamma)^{-1}$ the simulator reaches a steady state with distribution  $F_{\rm tls}$.
At this point the quantum simulator reaches equilibrium with the TLS environment at the temperature $T$ provided by the environment.
We note that for the steady state we can define $G^K(\omega)$ for which the fluctuation-dissipation theorem \eqref{eq:gk_decay} holds.
For $t\geq t'$ and $[\hat H_f,\hat \Gamma]_-=0$ we find the analytical solution
\begin{equation}
	G^K(t,t') = -i e^{-i \hat H_f t-\frac{\hat\Gamma}{2} t}
	\left[e^{\hat\Gamma t'}-2f(\hat H_{\rm ini})\right]e^{i \hat H_f t'-\frac{\hat\Gamma}{2}t'} \,.
\end{equation}
\begin{figure}[tb]
	\includegraphics[width=\columnwidth]{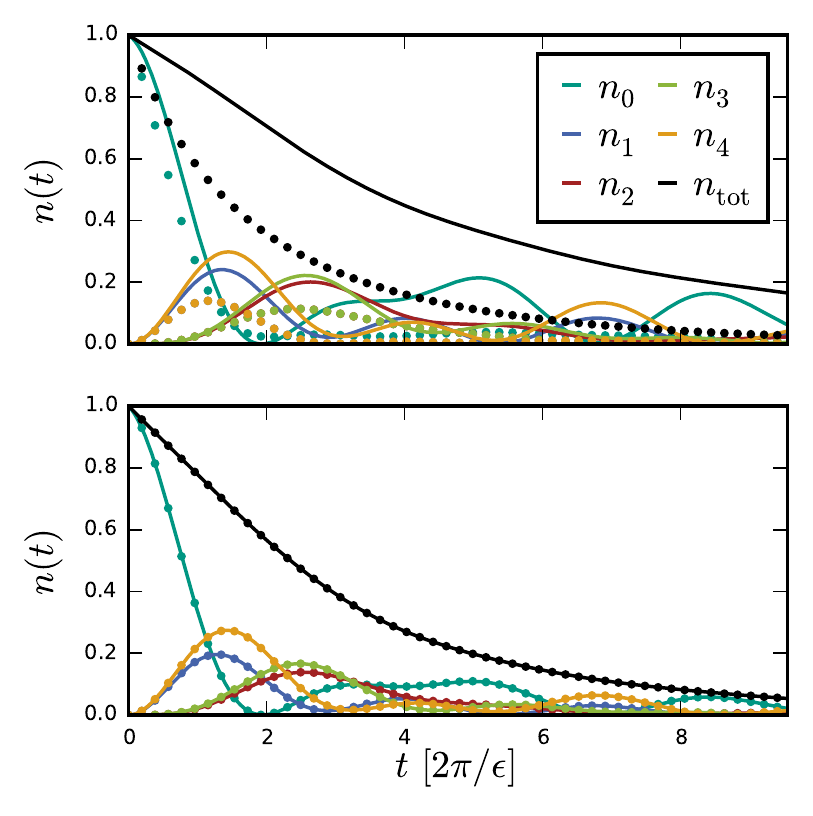}
	\vspace{-.75cm}
	\caption{\label{fig:time_dependent_decay}Numerical simulation of a tight binding chain with $N=5$ sites coupled to bath of TLS in the wide band approximation (bottom) and for decay due to a bath of 10 TLS in the ground state (top).
	 We plot the occupation $n_i(t)=\langle c_i^\dagger(t)c_i(t)\rangle$ as well as the total number of particles $n_{\rm tot}(t)=\sum_i n_i(t)$ in the system.
	 Initially qubit $i=0$ was in the excited state, all other qubits in the ground state. 
	 Solid lines are obtained via fermionic Keldysh theory and dots via quantum master equation.}
\end{figure}
If both time arguments are large compared to the smallest decay rate, $t,t'\gg \Gamma_{\rm min}^{-1}$, only the first term in the square brackets survives and the Green's function does no longer depend on the initial state. We find the stationary state solution $G^K = -i\exp\{(-i \hat h -\frac{\hat\Gamma}{2})(t-t')\}$ which depends only on the time difference and corresponds to a stationary solution with distribution function $1-2f=1$. This behavior resembles the expected behavior of a system in an arbitrary initial state subject to decay into an environmental bath. During a time $\sim \Gamma^{-1}$ the information of the initial state decays into the environment and the Keldysh Green's function depends only on the difference of time arguments. 
The decay of the initial state/ distribution function into an effective distribution with $\beta\to\infty$ describes the decay of particles into the bath of TLS at zero temperature.

To demonstrate the validity of the method and compare it to master-equation calculations, we solve the KBE numerically for a linear chain of five qubits coupled to a bath of TLS.
Initially, the zeroth qubit was excited, i.e.~$n_0(0)=\langle c_0^\dagger(0)c_0(0)\rangle=1$, while all remaining qubits were in the ground state. This corresponds to an initial density matrix with Hamiltonian $H_{\rm ini} = -c_0^\dagger c_0+\sum_{i=1}^{N-1}c_i^\dagger c_i$.
In Fig.~\ref{fig:time_dependent_decay} we plot the time dependent occupation numbers $n_i(t) = \frac12[G^K_{ii}(t,t)+i]$, and the total number of particles $n_{\rm tot}(t)$ in the system. 
We clearly see the expected exponential decay of the total number of particles with a rate dominated by the decay rate of the most populated qubit.
As expected for the wide-band limit, master equation and KBE yield the same results (bottom), while for a bath consisting of ten individual TLS with corresponding spectral density, KBE and master equation differ (top).

\section{Conclusions}\label{sec:Conclusions}
We have applied fermionic
non-equilibrium Green's function methods to a noisy analog quantum simulator. 
We used the Jordan-Wigner transformation to map the noisy qubit system to a coupled fermion-fermion (relaxation) and fermion-boson (dephasing) system.
Dephasing maps to a simple on-site interaction, $\sim c_i^\dagger c_i^{\phantom\dagger}$, and is equivalent to an electron-phonon interaction.
Using this connection we calculated disturbed Green's functions of noisy systems with diagrammatic many-body methods.
For small systems we compared the results with master-equation calculations. In the correct limit, both methods agree with each other.
We found that the spectral resolution of large quantum simulators is limited by the spectral width of the qubits.

Contrary to dephasing, the Jordan-Wigner mapping to fermionic degrees of freedom becomes troublesome for transverse coupling, $\propto\sigma_x$,
due to the non-conserved fermion number. In this work, we discussed the special case of a bath of TLS at low temperatures, which circumvented these problems.

The mapping to fermionic degrees of freedom comes with the advantage of the advanced and powerful many-body toolbox,
ranging from diagrammatic expansions to full field-theoretical approaches.
These methods help to establish a connection between the ideal Green's function that should be the result of quantum simulations and the perturbed Green's function obtained from a noisy quantum simulator.
\bibliography{bibliothek}
\addcontentsline{toc}{chapter}{\bibname}
\end{document}